\documentclass[%
 reprint,
 superscriptaddress,
 preprintnumbers,
 nofootinbib,
 amsmath,amssymb,
 aps,
 prl,
]{revtex4-1}
\usepackage{psfrag,slashed,cancel,array,graphicx,todonotes,hyperref}
\usepackage{subfigure}
\usepackage[labelfont=bf,labelsep=quad,justification=justified]{caption}
\usepackage[utf8]{inputenc}
\usepackage{mathtools}
\usepackage{mathrsfs}
\usepackage{multirow}
\usepackage{braket}
\usepackage{slashed}
\usepackage{booktabs} 
\usepackage[normalem]{ulem}
\usepackage{slashed}
\hypersetup{pdftitle={},pdfcreator={},linkcolor=[rgb]{0.15,0.35,0.75},colorlinks=true,citecolor=[rgb]{0.675,0,0.2},urlcolor=[rgb]{0.15,0.35,0.65}}

\def\beq{\begin{equation}}
\def\eeq{\end{equation}}
\def\bsp#1\esp{\begin{split}#1\end{split}}

\newcommand{\eq}[1]{eq.~\eqref{eq:#1}}

\newcommand{\cN}{\mathcal{N}}
\newcommand{\cO}{\mathcal{O}}

\newcommand{\nn}{\nonumber}

\newcommand{\df}{\mathrm{d}}

\newcommand{\Sl}[1]{\slashed{#1}}

\newcommand{\R}{R}
\newcommand{\nslash}{\Sl{n}}

\newcommand{\orange}[1]{{\color{orange}{#1}}}

\AtBeginDocument{
\heavyrulewidth=.08em
\lightrulewidth=.05em
\cmidrulewidth=.03em
\belowrulesep=.65ex
\belowbottomsep=0pt
\aboverulesep=.4ex
\abovetopsep=0pt
\cmidrulesep=\doublerulesep
\cmidrulekern=.5em
\defaultaddspace=.5em
}

\def\cP{\mathcal{P}}
\def\cO{\mathcal{O}}
\def\cC{\mathcal{C}}
\def\cN{\mathcal{N}}

\def\be{\begin{equation}}
\def\ee{\end{equation}}

\begin{document}

\preprint{MIT-CTP 5926, CERN-TH-2025-104}

\author{Hao Chen}
\email{hao\_chen@mit.edu}
\affiliation{Center for Theoretical Physics - a Leinweber Institute, Massachusetts Institute of Technology, Cambridge, MA 02139, USA}
\author{Pier Francesco Monni}
\email{pier.monni@cern.ch}
\affiliation{CERN, Theoretical Physics Department, CH-1211 Geneva 23, Switzerland}
\author{Zhaoyan Pang}
\email{zypang@stu.pku.edu.cn}
\affiliation{School of Physics, Peking University, Beijing 100871, China}\author{Gherardo Vita}
\email{gherardo.vita@cern.ch}
\affiliation{CERN, Theoretical Physics Department, CH-1211 Geneva 23, Switzerland}
\author{Hua Xing Zhu}
\email{zhuhx@pku.edu.cn}
\affiliation{School of Physics, Peking University, Beijing 100871, China}
\affiliation{Center for High Energy Physics, Peking University, Beijing 100871, China}

\title{Correlation Function/Wilson Loop Duality in Gauge Theory from EFT}

\begin{abstract}
In this Letter, we initiate a systematic study of the $n$-point correlation functions (CF) in gauge theories in the sequential light-cone (SLC) limit. Focusing on QCD, we formulate a factorization theorem for the CF of four vector currents in this limit using tools from soft-collinear effective field theory (SCET).
This result unveils a duality between CF and Wilson loops in non-conformal field theories, according to which the singular structure of the CF is described by a null polygonal Wilson loop dressed by universal jet functions and current form factors, directly related to well-known ingredients in collider physics.
We employ this new factorization theorem to obtain the singular terms of the four-point CF in QCD in the SLC limit up to three loops. This constitutes the first determination of terms beyond one loop and in particular determines, for the first time, conformal-symmetry-breaking terms induced by the non-vanishing $\beta$-function of QCD. As a stringent check, we verify that our result satisfies the anomalous conformal Ward identities and its leading-transcendental term reproduces the known correlation function in $\cN=4$ SYM to three loops.
Our results are easily generalized to the multi-point case, offering a powerful tool for future applications of SCET to the study of the fundamental structure of gauge theories.
\end{abstract}

\maketitle

\paragraph*{Introduction.---}

Correlation functions (CFs) are fundamental quantities in quantum field theory (QFT). They provide critical insight into the structural and dynamical properties of the underlying theory across high-energy physics, condensed matter systems, and cosmology. 
In the context of collider physics, correlation functions play a key role due to their relation of scattering amplitudes and cross sections~\cite{LSZ,Appelquist:1973uz,Zee:1973sr} as well as in the description of an important class of collider observable, that of collider correlators, that admit a representation in terms of these objects~\cite{Belitsky:2013xxa,Belitsky:2013bja,Belitsky:2013ofa}.
Besides being fascinating from a formal point of view, this representation is a powerful tool to obtain precise theory predictions for these observables by completely avoiding the complication of IR divergences cancellation in the more standard amplitude approach. This critically provides an alternative perspective on the factorization and renormalization group (RG) structure of collider observables, complementary to the standard diagrammatic and effective field theory approaches, which opens new angles to the study of collider physics with CFs~\cite{Hofman:2008ar,Belitsky:2013bja,Belitsky:2013ofa,Belitsky:2013xxa,Henn:2019gkr,Korchemsky:2019nzm,Kologlu:2019mfz,Chen:2020vvp,Korchemsky:2021okt,Lee:2022uwt,Lee:2023npz,Chen:2023wah,Chen:2023zzh,Monni:2025zyv}.

Although extensively studied in formal QFT, there are only a few concrete calculations of higher-point CFs of local operators in QCD~\cite{Chicherin:2020azt}. 
The non-conformal nature of QCD induces a more complicated kinematic dependence in local CFs and hence makes the calculations beyond one loop prohibitively difficult. Given the importance of CFs in the study of QCD, it is paramount to explore alternative methods for the quantitative analysis of these quantities. 
In this work, we initiate a study of higher-point local CFs beyond one loop in QCD by focusing on a special kinematic regime: the sequential light-cone (SLC) limit. 
This limit has a rich structure and encodes several interesting theoretical features. 
In $\cN=4$ SYM local CFs in the SLC limit are proportional to polygonal light-like Wilson loops~\cite{Alday:2010zy}, which in turn are interestingly dual to scattering amplitudes~\cite{Alday:2007hr,Drummond:2007aua,Brandhuber:2007yx,Drummond:2007cf,Drummond:2007au,Caron-Huot:2010ryg,Henn:2009bd,Chicherin:2022bov,Chicherin:2022zxo,Carrolo:2025pue}. There has also been interests in the geometric structure of such objects in terms of correlahedron and positive geometry~\cite{Eden:2017fow,He:2024xed,Arkani-Hamed:2017mur,Chicherin:2024hes,Brown:2025plq}.
They have also been systematically studied in conformal field theories via the light-cone bootstrap~\cite{Fitzpatrick:2012yx,Komargodski:2012ek,Alday:2013cwa,Alday:2015eya,Alday:2015ota,Alday:2015ewa,Alday:2016njk,Simmons-Duffin:2016wlq,Alday:2016njk,Bercini:2020msp,Bercini:2021jti} and integrability-based method~\cite{Belitsky:2019fan,Olivucci:2022aza}, which revealed many analytic features of large spin physics in these theories. 
Furthermore, in the context of collider physics, the SLC limit of four-point CFs is dual to the di-jet (back-to-back) limit of the aforementioned collider correlators~\cite{Korchemsky:2019nzm,Chen:2023wah}.
Remarkably, we find that, much like in conformal field theories, multi-point correlation functions in QCD are proportional to closed light-like Wilson loops in the SLC limit. 
The details of this connection are unveiled by means of a novel factorization formula that describes the CFs in this limit, and can be used to make quantitative predictions of their perturbative structure.
To our knowledge this is the first proof of such duality in any non-conformal gauge theory, and offers a powerful tool for the study of these key objects in an array of applications to collider physics and formal QFT aspects.

We consider the correlation function of four operators  
\be~\label{eq:JJJJ}
   G^{\mu\nu\rho\sigma}(\{x\}) =   \langle j^\mu(x_1) j^\nu(x_2) j^\rho(x_3) j^\sigma(x_4) \rangle\,,
\ee
where $j^\mu(x_i) = \bar{\psi}(x_i)\gamma^\mu\psi(x_i)$ are electromagnetic currents evaluated at points $x_i^\mu$ such that $y_i^2\equiv (x_{i+1}-x_i)^2 < 0$ are spacelike separated (with $x^\mu_5\equiv x^\mu_1$), and $\langle \cdot \rangle$ denotes a Wightman expectation value in the interacting vacuum. The region in which some or all of the $y_i^2 > 0$ is accessible through analytic continuation.
Note that the CF in Eq.~\eqref{eq:JJJJ} has a corresponding collider correlator, the charge-charge correlator (QQC) \cite{Hofman:2008ar,Belitsky:2013bja,Belitsky:2013ofa,Belitsky:2013xxa,Chicherin:2020azt}, which measures correlations between electromagnetic charges in electron-positron annihilation. 
A recent analysis~\cite{Monni:2025zyv} established that the QQC is infrared and collinear safe in the back-to-back limit and derived a corresponding factorization formula, enabling its perturbative resummation to N$^4$LL accuracy. 

\begin{figure}
  \centering
  \includegraphics[width=0.35\textwidth]{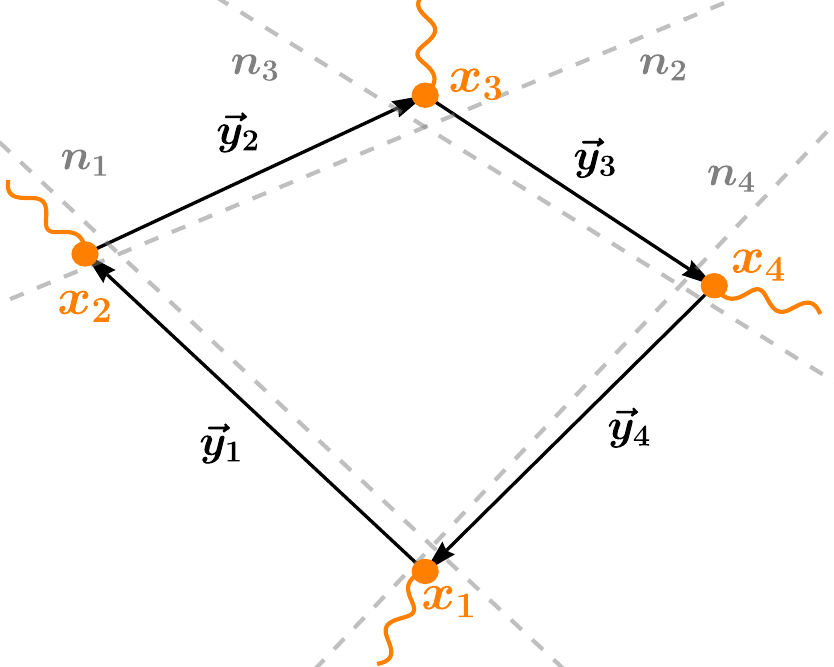} 
  \caption{Graphical depiction of $G(\{x\})$ in eq.~\eqref{eq:JJJJ} in the sequential light-cone limit. The dashed gray lines identify the light-cone directions approached by $y_i^\mu$.}
  \label{fig:setup}
\end{figure}

We are interested in the behavior of~\eqref{eq:JJJJ} in the SLC limit, that is where the $x_i^\mu$ are distinct and separated by a null distance, i.e. $y_{i}^2\to 0$. In this limit, $G$ exhibits pole singularities in $y_i^2$ related to the propagation of particles along the $y^\mu_{i}$ directions. At lowest perturbative order one finds 
\be~\label{eq:GatLO}
G^{\mu\nu\rho\sigma}_{\rm tree} (\{x\}) \sim \frac{ \mathrm{tr} \left[ \gamma^\mu \slashed{y}_1 \gamma^\nu \slashed{y}_2 \gamma^\rho \slashed{y}_3 \gamma^\sigma \slashed{y}_4 \right]}{(y_{1}^2)^2 (y_{2}^2)^2 (y_{3}^2)^2 (y_{4}^2)^2}+\dots\,,
\ee
where $y_i^\mu = x_{i+1}^\mu - x_i^\mu$, and we only kept the connected, leading term in this limit and neglected less singular Wick contractions in the calculation of eq.~\eqref{eq:JJJJ}. 
This situation is depicted schematically in fig.~\ref{fig:setup}. 
To study eq.~\eqref{eq:JJJJ} at higher loops we define the quantity 
\be~\label{eq:R}
\R=\lim_{y_{i}^2\to 0} G^{\mu\nu\rho\sigma}(\{x\})/G^{\mu\nu\rho\sigma}_{\rm tree}(\{x\}) 
\ee
which packages the radiative corrections and contains logarithmic lightcone divergences at each order in perturbation theory.
These light-cone divergences manifest themselves as logarithms of the conformal cross ratios $u= y_1^2 y_3^2/x_{13}^2x_{24}^2$ and $v= y_4^2 y_2^2/x_{13}^2x_{24}^2$, where $x^2_{13}=(y_2^\mu+y_1^\mu)^2<0$,~$x^2_{24}=(y_3^\mu+y_2^\mu)^2<0$, which become small in the SLC limit. 
In the following we will show that such logarithmic singularities are related to the infrared structure of classes of collider observables, and they exhibit a remarkably simple factorizing structure at all perturbative orders.

\paragraph*{Factorization in the sequential light-cone limit.---}
The starting point to build a factorization theorem for $\R$ is the identification of the power counting that defines the SLC limit.
Translational invariance forces $\R$ to depend on differences of the vertices coordinates, in terms of which the SLC limit can be expressed as 
\be\label{eq:DLlimitgeneral}
	|y_i^2| \ll  |(y_i + y_{i+1})^2| \sim  |x^2_{13}| \sim  |x^2_{24}| \sim L^2 \,,
\ee
where $L^2$ is the scale of the large separations $x^2_{13}$,~$x^2_{24}$. Note that we assume that $x^2_{13}\sim x^2_{24}$ so our discussion will be valid in this geometric configuration.
The physical picture is given by four wide angle light-like directions $n_i^\mu$ that the separations $y_i^\mu$ are approaching in the SLC limit.
To formalize the limit, we define the four light-like vectors $\{n^\mu_i\}_{i=1}^4$ as 
\be~\label{eq:light-cone-dec}
    n_i^\mu \equiv \left(\frac{y^0_i}{|y^0_i|},\,\frac{\vec{y}_i}{|\vec{y}_i|}\right)	 \,,\qquad n_i^2 =0 \,.
\ee
This allows us to define the standard light-cone decomposition for the $y_i^\mu$ vectors
\be
	y_i^\mu = y^+_i \frac{n_{i-1}^\mu}{n_i \cdot n_{i-1}} + y_{i}^- \frac{n_{i}^\mu}{n_i \cdot n_{i-1}} + y_{i,\perp}^\mu \equiv (y^+_i,y_{i}^-,y_{i,\perp})
\ee
with $	y_i^2 = 2 y^+_{i} y^-_{i}/(n_i\cdot n_{i-1}) + y_{i,\perp}^2$, $y_{i,\perp}^2<0$. 
Eq.~\eqref{eq:DLlimitgeneral} implies the following hierarchy between the components of $y_i$ 
\begin{equation}
	\frac{y^+_{i}}{y_{i}^-} \sim \lambda^2\ll 1\,.
\end{equation}\label{eq:yihierarchies}
Moreover, in order for the factorization to capture the endpoint $y_i^2=0$ we also require $y^+_{i} y^-_{i} \sim y_{i,\perp}^2 $,
leading to the definition of a \textit{collinear} scaling for $y_i$
\be
	y_i^\mu \sim L (\lambda^2,1,\lambda)\,,\quad |y_i^2| \sim L^2 \lambda^2\,.
\ee

\begin{figure}
  \centering \includegraphics[width=0.37\textwidth]{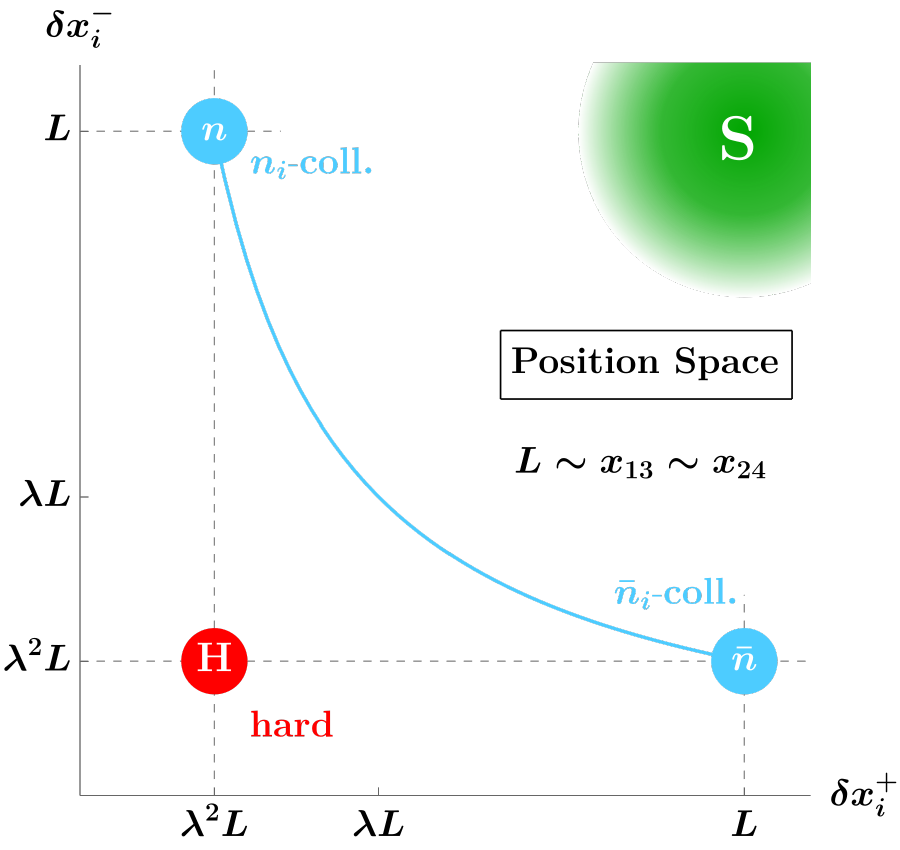}  \caption{SCET modes in position space. }
  \label{fig:modesxspace}
\end{figure}

Therefore, taking the SLC limit is formally equivalent to probing the $y_i^\mu$'s in configurations where the time and spatial components are large but very close in size, or equivalently, where one light-cone component in \eq{light-cone-dec} is much smaller than the other. To give a physical interpretation to this problem we can think of the momenta $q_i$ conjugate to the coordinates $y_i$. We then have that the $q_i^-$ are large components, related to the small scale fluctuations that determine the separation between $y_i^\mu$ and the light-like directions $n_i^\mu$.
The $q_i^+$ are instead much smaller, conjugate to the large separation present between $y_i^\mu$ and the other light-like directions $n_{j\neq i}^\mu$.
This crucially implies that we can think of the limit of $y_i$ to the light-cone as a measurement, in momentum space, of the projection of its conjugate momentum $q_i$ along a direction $n_i^\mu$. 
This analogy maps the SLC limit of $\R$ in \eq{R} to the infrared limit of well-known collider observables which is characterized by a large hierarchy of scales. Specifically, a measurement of this kind corresponds to the thrust event-shape distribution~\cite{Farhi:1977sg}.
Crucially, this provides us with a systematic way of constructing an effective field theory to describe the SLC limit, 
in line with the well-known SCET$_{I}$ theory~\cite{Bauer:2000ew, Bauer:2000yr, Bauer:2001ct, Bauer:2001yt}. 

We can derive a factorization theorem by employing standard techniques in SCET literature, see e.g.~\cite{Fleming:2007qr,Bauer:2008dt}. First, each electromagnetic current is matched to a non-local operator involving a pair of gauge invariant collinear fields
\be
    j^\mu(x_i) \to \!\! \int\!\! \df s\, \df t\,  \cC(s,t)  \bar{\chi}_{n_{i}}(x_i+s \,n_{i-1}) \gamma^\mu_\perp \chi_{n_{i-1}} (x_i+t\, n_{i})\,,
\ee

where $\cC$ is the finite part of the photon-quark form factor which is known up to 4 loops in QCD~\cite{Lee:2022nhh}. 
Each collinear field can be decoupled from soft interactions via an appropriate field redefinition \cite{Bauer:2001yt}, introducing soft Wilson lines which combine to give the expectation value of a null polygonal Wilson loop 
\be~\label{eq:wilson-loop}
   \mathcal{W}(\{x\}) = \langle \cP \exp\left\{ig\int_{\Gamma(\{x\})} \!\!\!\!\!\!\df z^\mu {A^a_s}_\mu(z) T^a\right\} \rangle\,.
\ee
Here $\Gamma(\{x\})$ is the path along the polygon composed by segments that exactly lie on the light-cone directions $n_i$ and have lengths equal to the large, $\cO(L)$, component of the $y_i$'s. 
Finally, the pairs of collinear fields belonging to the same direction are combined into \emph{jet functions}, well known SCET objects that are closely related to the propagator of these collinear fields. They read~
\begin{align}\label{LaplaceJet}
	\mathcal{J}_{n_i}&(\frac{\omega}{y_i^-},\mu)=-\frac{1}{\pi}\int_0^{\infty} \df p^2 e^{\frac{y_{i}^{-}}{2\omega} p^2} \\
    & \times \Im\left[\frac{i}{\omega}\int \df^4 x\, e^{-i p\cdot x}\langle T \bar{\chi}_{n_i}(x)\frac{\nslash_{i-1}}{4\,N_c}\chi_{n_i}(0)\rangle\right]\notag\,,
\end{align}
and can be extracted to three loops from the calculation of ref.~\cite{Bruser:2018rad}.

In the end, we arrive at the following remarkably simple factorization theorem for $\R$ at leading power in the SLC limit
\begin{equation}~\label{eq:fact}
\R = {\cal W}(\{x\}) \prod_{i=1}^4 \int_0^\infty \!\!\! \df \omega_i \, \cC_i(\omega_i \omega_{i+1}) {\cal J}_{n_i}(\frac{\omega_i}{y_i^-})  e^{-\omega_i \frac{y_i^2}{2 y_i^-}} \,.
\end{equation}

\begin{figure}
  \centering
  \includegraphics[width=0.4\textwidth]{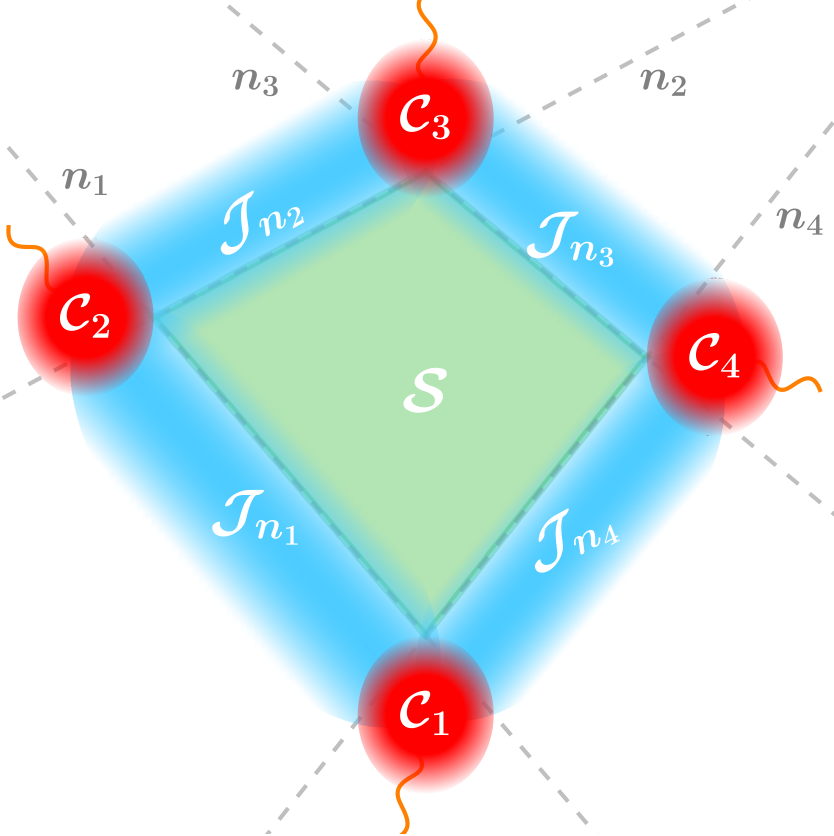} 
  \caption{Pictorial view of the factorization theorem for the sequential light-cone limit of the four-point correlation function. The Soft function, represented by the green polygon is a Wilson loop tracing the null polygon.}
  \label{fig:factthm}
\end{figure}

Eq.~\eqref{eq:fact} reveals that the CF in the SLC limit can be simply described by the product of a polygonal Wilson loop and convolutions of form factors with jet functions. Importantly, each of the elements of the factorization formula~\eqref{eq:fact} is a well known object in collider physics, entering the theoretical description of the infrared structure of collider observables such as event shapes. An important consequence of this observation is that both the finite remainder of the form factor $\cC_i$ and the jet function ${\cal J}_{n_i}$ can be extracted at high-loop orders from the literature~\cite{Bruser:2018rad,Lee:2022nhh}. On the other hand, the Wilson loop ${\cal W}(\{x\})$ in QCD was previously unknown beyond one loop, and we compute it to two loops in this Letter (all results are provided in ref.~\cite{zenodo_2025_17200473}).
The ingredients $\{{\cal W},\cC_i,{\cal J}_{n_i}\}$ of eq.~\eqref{eq:fact} also depend on a common renormalization scale $\mu$, that has been omitted in eq.~\eqref{eq:fact} to simplify the notation. This dependence is governed by the following linear RG equations that can be used to resum the large logarithmic corrections in $R$
\begin{align}
\begin{split}\label{eq:RGEs}
    \frac{\df \ln{\cal W}(\{x\})}{\df \ln \mu} &= -2\Gamma_{\mathrm{cusp}}\left(\alpha_s\right) \ln \frac{x_{13}^2 x_{24}^2 \mu^{4}}{16} + \gamma^W\left(\alpha_s\right) \,,\\
    \frac{\df \ln{\cal J}_{n_i}(\omega_i/y_i^-)}{\df \ln \mu} &= -2 \Gamma_{\mathrm{cusp}}\left(\alpha_s\right) \ln \frac{2\, \omega_{i}}{y_i^-\mu^2}+ \gamma^J\left(\alpha_s\right) \,,\\ 
    \frac{\df \ln\cC_i(\omega_i \omega_{i+1})}{\df \ln \mu} &=  \Gamma_{\mathrm{cusp}}\left(\alpha_s\right) \ln \frac{2\,\omega_i\omega_{i+1}}{\mu^2 n_i\cdot n_{i-1}} + \gamma^C\left(\alpha_s\right)\,,
\end{split}
\end{align}
where $x_{13}^2 = 2 y_1^- y_2^-/n_1\cdot n_4= 2 y_3^- y_4^-/n_2\cdot n_3$ and $x_{24}^2 = 2 y_1^- y_4^-/n_3\cdot n_4= 2 y_2^- y_3^-/n_1\cdot n_2$ up to power corrections in $\lambda$.
The $\mu$ dependence cancels order-by-order in the r.h.s. of eq.~\eqref{eq:fact} such that $R$ is RG invariant leading to the following consistency relation for the non-cusp anomalous dimensions 
\be
    \gamma^{W_n}(\alpha_s) = -n \gamma^J (\alpha_s) - n \gamma^C (\alpha_s)  = -\frac{n}{2} \gamma^{\text{thr.}} (\alpha_s)\,,
\ee
where $\gamma^{\text{thr.}} (\alpha_s)$ is the soft anomalous dimension for threshold resummation that is currently known to four loops~\cite{Das:2020adl,Duhr:2022cob,Moult:2022xzt}, analogously to $\gamma^J$~\cite{Das:2019btv,Duhr:2022cob} and $\gamma^C$~\cite{vonManteuffel:2020vjv}.
Our derivation of the factorization formula for the CF in the SLC limit can be easily extended to the $n$-point case. This simply entails evaluating the Wilson loop ${\cal W}$ on the $n$-gon $\Gamma$ defined by the $n$ light-like directions of the problem, and extending the product in eq.~\eqref{eq:fact} up to $n$. The resulting equation for the $n$-point CF constitutes the generalization of the results of ref.~\cite{Alday:2010zy} to non-conformal gauge theories such as QCD. In this case, our result reveals that even without the presence of a conformal symmetry the CF is dual to a polygonal Wilson loop. However, the proportionality factor has a more complex structure than in the conformal case, and it is entirely captured by our factorization formula.

\paragraph*{Predictions for $\cN=4$ SYM and QCD.---}
Eq.~\eqref{eq:fact} can be used to calculate the CF in the SLC limit in different QFTs. We examine the four-point case in QCD, for which we extract the three loop form factor and jet function from Refs.~\cite{Bruser:2018rad,Lee:2022nhh}. The four-point two-loop Wilson loop is computed for the first time in this Letter~\cite{zenodo_2025_17200473}, and we predict its three-loop structure up to terms that are regular in the SLC limit using the RG equations~\eqref{eq:RGEs}. We work in conventional dimensional regularization scheme in $D=4-2\epsilon$ dimensions, which allows us to verify the cancellation of all $\epsilon$ poles and the consistency of the (local) renormalization of the ingredients of eq.~\eqref{eq:fact}. 
We write the result as an expansion in the strong coupling $\alpha_S$ as $R = 1 + \sum_n\frac{\alpha^n_S(\mu)}{(4 \pi)^n} R_n$. The one-loop result is given by
\begin{equation}\label{eq:one-loop}
    R_1 = {\color{blue}C_F}  \left(-4 \zeta _2-2 L_u L_v + L_{uv} \right) \,,
\end{equation}
where we have introduced the abbreviation $L_u = \ln u$, $L_v = \ln v$, and $L_{uv} = \ln(u v)$. The one-loop result of Eq.~\eqref{eq:one-loop} can also be extracted by taking the SLC limit of the full one-loop result of Ref.~\cite{Chicherin:2020azt}.

Our factorization theorem \eq{fact} allows us to calculate the new two-loop result as well as all the logarithmic terms at three loops (while the calculation of regular terms in the SLC limit requires the knowledge of the four-point Wilson loop at three loops, which is currently unknown). These read
\begin{widetext}
{\footnotesize
\begin{align}
    R_2 = &\ {\color{blue}C_F^2} \Bigg[
       -28 \zeta _2-48 \zeta _3+116 \zeta _4+L_u \left(8-2
   L_{uv}\right) L_v+2 L_u^2 L_v^2
   +
   \left(4\zeta _2-\frac{7}{2}\right)
   L_{uv}^2+\left(4 \zeta _2-24 \zeta
   _3+\frac{5}{2}\right) L_{uv}  
   \Bigg]    \\& + {\color{blue} C_F C_A} \Bigg[
      \frac{68 \zeta _2}{9}
   +\frac{28 \zeta _3}{3}-16 \zeta
   _4+L_u L_v \left(4 \zeta _2-\frac{11
   L_{uv}}{6}-\frac{235}{18}\right) +\frac{11
   L_{uv}^2}{12}
    +\left(-\frac{112 \zeta _2}{3}+12 \zeta
   _3+\frac{221}{18}\right) L_{uv}
        \Bigg]
   \nn\\& + {\color{blue} C_F n_f}
    \Bigg[
       -\frac{8 \zeta _2}{9}+\frac{8 \zeta _3}{3} 
     +
    L_u \left(\frac{L_{uv}}{3}+\frac{17}{9}\right)
   L_v+\left(\frac{16 \zeta
   _2}{3}-\frac{13}{9}\right)
   L_{uv}-\frac{L_{uv}^2}{6}
    \Bigg]    
  + 16 {\color{blue} \left( C_F^2 - \frac{1}{2}C_F C_A\right) } \nn\\&
  + \frac{1}{2} \orange{\beta_0}  R_1 \ln \frac{x_{13}^2 x_{24}^2 \mu^4}{16}  
   + {\color{blue} C_F} \orange{\beta_0}  \Big[r_2(L_{uv}, x) + r_2\big(L_{uv}, \frac{1}{x}\big) 
+\frac{L_{13}^2}{4}+\frac{L_{24}^2}{4} 
+\ln (x)L_{13} L_{24}-\frac{1}{2}\left(L_v L_{13}^2+ L_u L_{24}^2\right)\Big],\nn
\end{align}
\par}

{\footnotesize
\begin{align}\label{eq:R3}
R_{3}=& 
C_{A}C_{F}^2\left(-\frac{4}{9} \pi^4 L_u^2-\frac{2}{45} \pi^4 L_u L_v-\frac{2}{3} \pi^2 L_u^2 L_v^2-\frac{56}{3} \pi^2 L_u \zeta_3 
-24 L_u^2 L_v \zeta_3-120 L_u \zeta_5\right) + C_{A}^2C_{F}\left(-\frac{11}{45} \pi^4 L_u L_v-40 L_u \zeta_5\right)\\
&  + C_{F}^3\left(-\frac{4}{9} \pi^4 L_u^2-\frac{26}{15} \pi^4 L_u L_v-\frac{4}{3} \pi^2 L_u^3 L_v-\frac{2}{3} \pi^2 L_u^2 L_v^2-\frac{2}{3} L_u^3 L_v^3 +32 \pi^2 L_u \zeta_3-\frac{16}{3} L_u^3 \zeta_3+48 L_u^2 L_v \zeta_3+240 L_u \zeta_5\right) + \left[ u \leftrightarrow v\right]+\dots\,\nn
\end{align}
}
\end{widetext}
\noindent We defined $\beta_0 = 11 C_A/3 - 2 n_f/3$, $x = x_{13}^2/x_{24}^2$, $L_{13}=\ln ({y_{1}^2}/{y_{3}^2})$, $L_{24}=\ln ({y_{2}^2}/{y_{4}^2})$ and
\begin{align}
    r_2(L_{uv}, x) =&\ 2 \text{Li}_3(-x) - 2 \ln(x) \text{Li}_2 (-x) - \frac{1}{4} L_{uv} \ln^2(x)
    \nonumber\\
    &\hspace{-1.5cm}\ - \frac{1}{2} (6 \zeta_2 + \ln^2(x) ) \ln\frac{(1+x)^2}{x} +\frac{1}{4}\ln^2 (x)\,.
\end{align}%

\paragraph*{Checks and discussion.---}
A first consistency check of the validity of the factorization theorem in \eq{fact} is the complete cancellation of poles in our perturbative results. This check is far from trivial as the cancellation happens after combining all the bare ingredients via the convolution structure dictated by \eq{fact}. Moreover, both the hard Wilson coefficient and jet functions are fixed and taken from independent results, while the soft function directly follows from the calculation of the closed Wilson loop. On a related note we have also independently checked that, after individually renormalizing the objects in the factorization theorem, the scale dependence of the renormalized objects cancels when combined according to \eq{fact}. In other words this implies that the factorization theorem holds both at the bare and at the renormalized level, further strengthening its validity. 
As a consistency check of our results, we verify that our expressions for the CF satisfy the anomalous conformal Ward identities given in ref.~\cite{Braun:2003rp}, expanded in the sequential lightlike limit
\begin{align}\label{eq:CWI}
    K^\mu \langle j_1 \dots j_4 \rangle = -\beta(g)\! \int\! d^4x 2x^\mu \langle \frac{\partial}{\partial g} \mathcal{L}(x) j_1 \dots j_4 \rangle\,,
\end{align}
where $j_i\equiv j_{\mu_i}(x_i)$, $\beta(g)$ is the QCD beta function, $\mathcal{L}$ is the QCD Lagrangian, and $K^\mu$ is the special conformal transformation
\begin{equation}
    K^\mu = \sum_{i=1}^4 2x_i^\mu(3+x_i\cdot \partial_{x_i})-2\Sigma^\mu_{\,\nu} x_i^\nu -x_i^2 \partial^\mu_{x_i}\,.
\end{equation}
We have verified with an explicit calculation that the action of $K^\mu$ on our two-loop prediction for the correlation function agrees with the r.h.s. of eq.~\eqref{eq:CWI}, that we computed following ref.~\cite{Chicherin:2020azt} by evaluating the position-space integrals with a Lagrangian insertion. 
As a further check, we use the principle of leading transcendentality to obtain the two- and three-loop results for ${\cal N}  =4$ SYM by performing the replacements $C_F \to N_c$ and $C_A \to N_c$, and keeping only the leading transcendental terms in the QCD expressions given above. For instance, for the three-loop correction we find
\begin{equation}
\begin{aligned}
 R_3^{{\cal N} = 4} =& -N_{c}^3\left[ \frac{2}{3} L_u^3 L_v^3+\frac{4}{3} \pi^2 L_u^2 L_v^2+\frac{4}{3} \pi^2 L_u^3 L_v\right.\\
&+\frac{16}{3} L_u^3 \zeta_3- 24 L_u^2 L_v \zeta_3 +\frac{8}{9} \pi^4 L_u^2+\frac{91}{45} \pi^4 L_u L_v \\
&\left. -\frac{40}{3} \pi^2 L_u \zeta_3- 80 L_u \zeta_5 \right] + \left[ u \leftrightarrow v\right]\,,
\end{aligned}
\end{equation}
which is in full agreement with the direct calculation performed in ref.~\cite{Drummond:2013nda}. This provides a non-trivial check of our procedure.

\paragraph*{Conclusions.---}
In this Letter we have initiated a systematic study of the sequential light-cone limit of multi-point correlation functions in QCD. We have showed that this limit admits an effective-field-theory description owing to the large hierarchy of scales in the problem. We showed how SCET can be used to derive a factorization formula for the multi-point CFs that generalizes the duality between these objects and Wilson loops, previously observed for conformal field theories, to non-conformal gauge theories such as QCD. This result provides a powerful tool to predict the perturbative structure of these key objects in different QFTs, that we exploit to derive the first prediction for the full singular behaviour of the correlator of four electro-magnetic currents in the SLC limit at two loops in QCD, and of its log-enhanced terms at three loops.

This work opens several interesting avenues that are worth exploring in the future. A first line of research concerns the study of the multi-point case, which will require the calculation of the finite terms of the $n$-gon Wilson loop with more than four cusped vertices. Secondly, it will be interesting to apply the techniques developed in this Letter to study other correlation functions, for instance correlators of the electromagnetic current and the energy-momentum tensor, which enter the description of energy correlators at collider experiments. A third, related direction concerns the use of the results presented in this article to predict the structure of collider correlators, such as the back-to-back limit of the charge-charge correlation~\cite{Monni:2025zyv}. This crucially requires understanding the details of the detector limit and light transform of the renormalized CF which is still an open problem in non-conformal   theories.
The above explorations will greatly deepen our knowledge of CFs in QCD and their connection with scattering observables, offering a novel angle to approach the derivation of precise theoretical predictions at collider experiments.

\paragraph*{Acknowledgements.---}
 We thank A.~Zhiboedov for valuable discussions and comments on the draft, and D.~Chicherin for helpful discussions and correspondence.
 We also thank the participants of Resummation, Amplitudes, and Subtractions 2024 and 2025 at CERN, and the Energy Operators in Particle Physics, Quantum Field Theory and Gravity at Simons Center for Geometry and Physics for useful discussion.
 PM and GV wish to thank the Center for High Energy Physics of Peking University for hospitality while this work was carried out.
 HC is supported by the U.S. Department of Energy, Office of Science, Office of Nuclear Physics under grant Contract Number DESC0011090.
 ZYP and HXZ are funded by the National Natural Science Foundation of China under contract No.~1242550.
 The work of PM and GV is funded by the European Union (ERC, grant agreement No. 101044599). Views and opinions expressed are however those of the authors only and do not necessarily reflect those of the European Union or the European Research Council Executive Agency. Neither the European Union nor the granting authority can be held responsible for them.
 
\bibliographystyle{apsrev4-2}
\bibliography{refs}

\end{document}